# Challenges of Adopting SAFe in the Banking Industry – A Study Two Years after its Introduction

Sara Nilsson Tengstrand[1], Piotr Tomaszewski[2(✉)][0000-0001-7877-2121], Markus Borg[1,2][0000-0001-7879-4371], Ronald Jabangwe[2][0000-0002-2305-6352]

[1] Lund University, Lund, Sweden
[2] RISE Research Institutes of Sweden, Lund, Sweden
{piotr.tomaszewski,markus.borg,ronald.jabangwe}@ri.se

**Abstract.** The Scaled Agile Framework (SAFe) is a framework for scaling agile methods in large organizations. We have found several experience reports and white papers describing SAFe adoptions in different banks, which indicates that SAFe is being used in the banking industry. However, there is a lack of academic publications on the topic, the banking industry is missing in the scientific reports analyzing SAFe transformations. To fill this gap, we present a study on the main challenges with a SAFe transformation at a large full-service bank. We identify the challenges in the bank under study and compare the findings with experience reports from other banks, as well as with research on SAFe transformations in other domains. Many of the challenges reported in this paper overlap with the generic SAFe challenges, including management and organization, education and training, culture and mindset, requirements engineering, quality assurance, and systems architecture. However, we also report some novel challenges specific to the banking domain, e.g., the risk of jeopardizing customer relations, stability, and trust of external stakeholders. This study validates several SAFe-related challenges reported in previous work in the banking context. It also brings up some novel challenges specific to the banking industry. Therefore, we believe our results are particularly useful to practitioners responsible for SAFe transformations at other banks.

## 1    Introduction

Even though agile methods have become popular among all kinds of companies, the methods were initially created for small teams and organizations. This often causes challenges when large organizations want to go agile [1]. Several frameworks guide a large-scale agile adoption, such as Large Scale Scrum, Disciplined Agile Delivery, and the Scaled Agile Framework (SAFe) [2]. According to a yearly survey by VersionOne, a company specializing in agile solutions, SAFe is the most popular framework for scaling agile in large enterprises [3]. SAFe is a set of principles and practices that aims at scaling agile methods for large organizations.

Scaled Agile, the company behind SAFe, publishes experience reports from companies that have introduced SAFe. The financial sector, and the banking industry specifically, are represented there [2]. At the time of writing, several actors from the banking industry are in the middle of a SAFe-transformation. Scaled Agile quotes experience reports from SAFe introductions at banks such as Nordea, Standard Bank, and Capital One [4], showing that SAFe is actively used in the banking industry.

As it can be expected, there are numerous challenges for both large-scale agile transformations in general and SAFe transformations specifically [1][2][5]. Despite that fact, Putta et al. [2] report a lack of scientific research on the challenges of SAFe adoptions in general. The banking industry is no exception in this matter.

Our study aims to fill that gap by identifying the challenges for a SAFe transformation in the banking industry. We seek an answer to the following research question:

> *RQ*: *What are the main challenges for adopting SAFe in the banking industry?*

To answer this question, we conduct a qualitative survey at a large full-service bank. The study consists of several interviews with people representing key roles involved in the SAFe transformation. To establish how our findings fit into the existing body of knowledge and how much they can be generalized, we compare them to the existing research on challenges with general agile transformations and to the aforementioned SAFe introduction experience reports.

The rest of the paper is organized as follows: Section II introduces theory on large-scale agile software development and the Scaled Agile Framework. Section III presents related work and Section VI describes the research method. Section V presents the results of the study; Section VI discusses the results. Section VII concludes the findings.

## 2      Large Scale Agile Software Development

Dikert et al. [1] define large-scale agile software development as software development organizations with 50 or more people or at least six teams. Agile methods were originally created with small and isolated teams in mind. Scaling up agile practices to larger organizations with multiple teams poses certain difficulties. Software development in large companies often means larger projects that span over a long period. Moreover, there is often a need to coordinate multiple teams in the software development process. To be able to meet the needs of large organizations, agile practices need to be applied to the entire organization. There have been a number of scaled agile frameworks proposed and, as mentioned, the most popular amongst them is SAFe [3].

SAFe is a set of principles and practices that make it possible to apply an agile way of working throughout the entire organization. SAFe can be configured in different ways and can be adapted to the specific needs of the company. The framework is built on core values and principles and has an implementation roadmap to guide organizations on how to go through the transformation. SAFe is suitable for companies ranging from medium-sized, with roughly 50 employees, to large with thousands of people [6].

SAFe offers several different configurations. In the most extensive configuration, there are guidelines for four defined levels of the organization: team, program, value stream/solution, and portfolio. Figure 1 shows some examples of roles and activities. At the Team level, the Scrum master, agile teams, and the product owner are operating and delivering working systems at least every two weeks. The development is primarily based on user stories and enabler stories. At the Program level, the agile teams are coordinated by an Agile Release Train (ART). ART consists usually of five to twelve teams that work together coordinated by the Release Train Engineer. This level focuses on creating artifacts such as a vision, roadmaps, and features. The Value stream level, sometimes called the solution level, is for organizations that require additional roles to integrate the work of complex systems that are dependent on each other. At this level, release management roles work with economic frameworks to coordinate multiple ARTs and value streams. The Portfolio level has the purpose of aligning the value streams from the lower levels to meet the business goals and financial goals of both the portfolio and the organization's overall business goals by program portfolio management. At this level, there are so-called Epics, which are initiatives that transcend all levels of the organization, i.e., from the visions of the upper levels to concrete development projects in the lower levels [7].

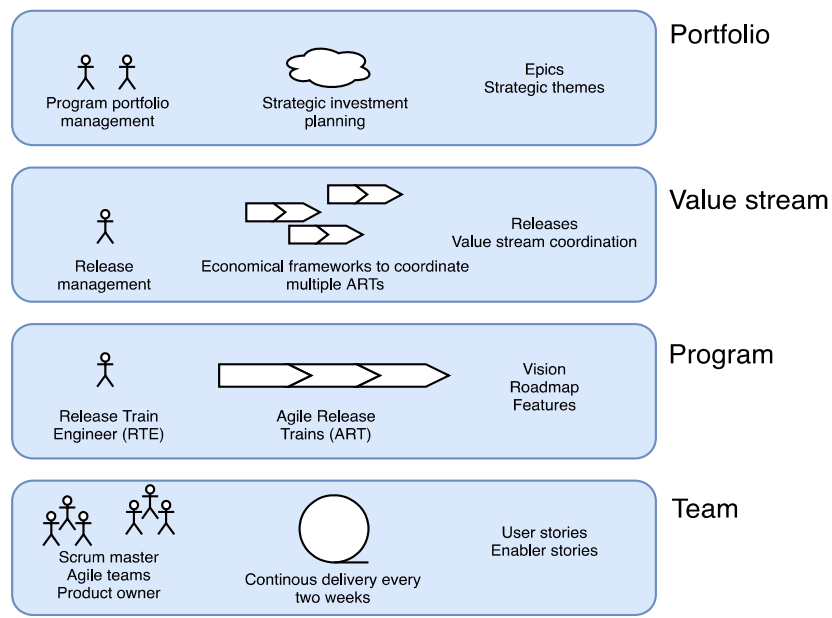

**Fig. 1.** Examples of roles and activities included at each level of SAFe

## 3  Related Work

This section presents individual studies on SAFe transformations from the literature, followed by three secondary studies.

### 3.1 Individual Case Studies

Although there is little research on SAFe transformations in the banking industry, there are studies on large-scale agile transformations in other industries. Maarit et al. [8] have made a case study on the organization-wide transformation at Nokia and found that the main group of challenges is related to the deployment of agile methods and the second largest group of challenges is managing and planning requirements according to agile methods. Paasivaara et al. [9] have made a case study on Ericsson's large-scale agile transformation and experienced challenges such as change resistance, lack of training and coaching, and that the surrounding organization was still in "waterfall mode". Paasivaara et al. [10] have further made a case study on the SAFe transformation at Comptel, a globally distributed software development company. In the case of Comptel, challenges such as lack of early training sessions and change resistance are reported.

Several white papers reporting challenges from SAFe transformations in full-service banks are available on the Scaled Agile website. Nordea introduced SAFe in 2014. One challenge that was reported is that different teams in the same Agile Release Train were frustrated because the delivery streams were not in unison, meaning it was not clear what was supposed to be delivered. However, the so-called Program Increment (PI) sessions, which is a planning event preceding the PI, which in its turn is a 10-week period consisting of five sprints, were reported as a successful way of uniting the teams [11]. Initio, a business consultancy firm, has analyzed the effect of implementing a SAFe transformation at ING Benelux, BNP Paribas, Deutsche Bank, and SimCorp and reports lessons learned from the transformations. Observations include that changes need to happen incrementally by doing small experiments frequently. Moreover, they recommend a close collaboration with senior management and training senior management in SAFe principles and practices [12]. The South African Standard Bank reveals some challenges when transforming their company according to SAFe. Standard Bank rolled out a few agile teams but experienced difficulties when scaling up the agile methodologies and having teams working together [13]. Johnston & Gill [14] have further investigated the case of Standard Bank and found that it was challenging to redefine the project manager role, specifically replacing the command-and-control leadership style with a coaching one. Furthermore, it was difficult for higher management to understand the long term benefits of the transformation. Making tradeoffs between quality and time as well as requirements prioritization have also been listed as challenges. The American Bank Capital One reports that it was hard for teams to accept the change and that early on in the transformation, it was difficult for teams to deliver independently because of dependencies outside of the teams [15].

Berkani et al. conducted a case study on an agile transformation in a French central bank [16]. The study is motivated by the research gap on how a company goes from experimenting with agile methods to establishing agile methods as a natural part of the organization. The study lists factors for a successful implementation in a large organization such as a reorganization of the Project Management Office and IT projects department, and generalizations of internal agile methods from a top management perspective. However, the study does not describe challenges related to the transformation.

## 3.2 Systematic Literature Reviews

We have identified three different systematic literature reviews that summarize and categorize findings from studies on large scale agile transformations [1][5][2]. Figure 2 summarizes the categories of challenges found in these studies. The first two studies focus on large-scale agile transformations in general, while the third study focuses on SAFe transformation in particular. In the first study by Dikert et al., the category where most challenges are reported is "agile difficult to implement" and an example of a challenge in that category is that agile is customizable poorly [1]. The second study by Uludag et al. reports most challenges with coordinating multiple agile teams that work on the same product, which is a challenge in the "communication and coordination" category [5]. In the third study by Putta et al., which focuses specifically on SAFe transformations, the most common challenge is change resistance, found in the "organizational and cultural" category [2].

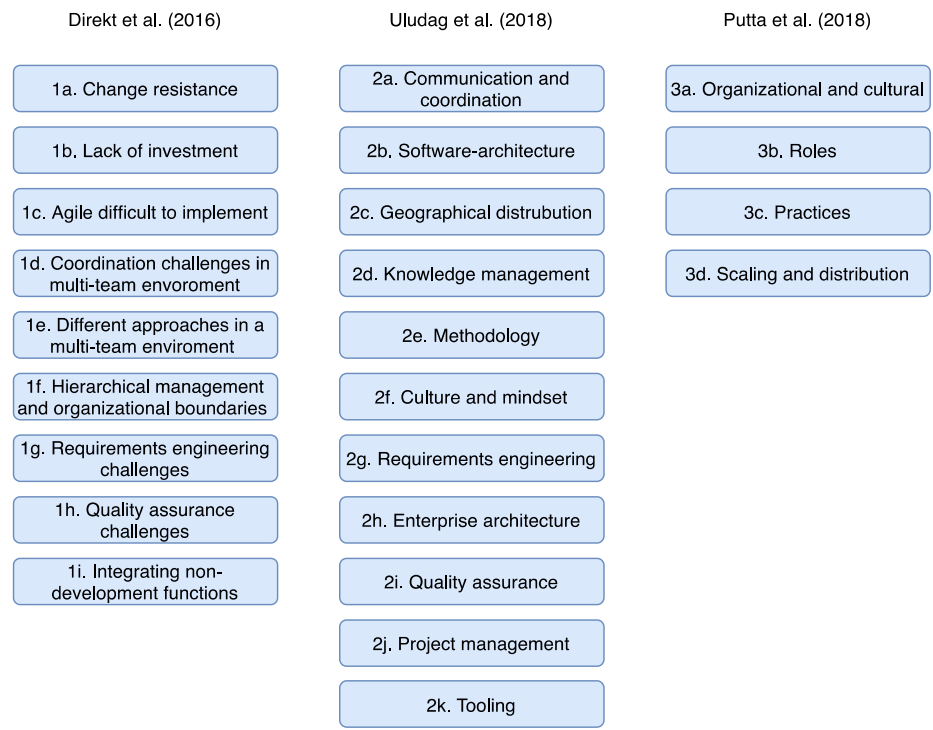

**Fig. 2.** Categories of challenges presented in three different systematic literature reviews. Note that the ordering within each column does not convey any particular meaning.

## 4    Method

Figure 3 shows the research strategy. The study is exploratory since it seeks new insights into the SAFe transformation challenges in the case company and the banking industry in general.

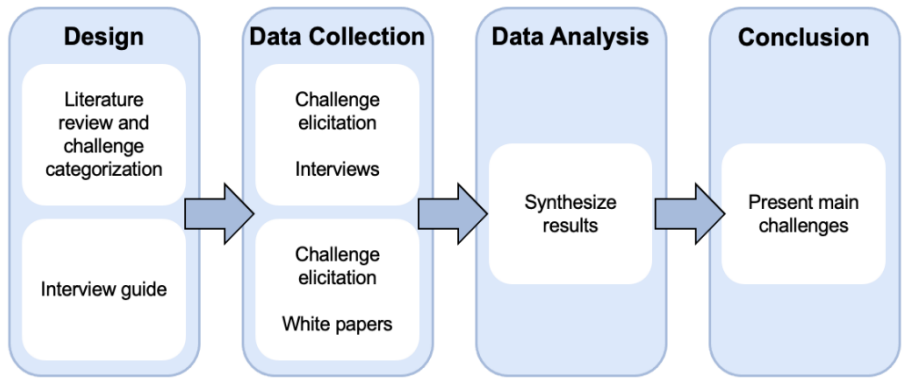

**Fig. 3.**    The research strategy for the study

In order to establish the current body of knowledge with regards to the agile transformations and challenges associated with them we have performed a literature review. The findings from the literature review, primarily from the existing systematic literature reviews, formed a base for an interview study that we conducted in a large bank. The purpose of the study was to identify the challenges the bank faced during the SAFe transformation. On top of that we have looked into a number of white papers where other banks described their own experiences from agile transformations. Finally, we have compared and synthesized the results.

The company where the interviews were performed is a large full service bank with services such as consumer banking, investment banking and trading. Examples of consumer banking services are customer saving accounts and mortgages whereas investment banking includes services such as assisting companies in mergers and acquisitions. The bank has physical branches where they offer face-to-face services to their customers as well as online presence. Furthermore, the case company operates in several countries and has over 12,000 employees globally. The bank has a large IT department that is responsible for customer facing products, like internet banking, but also products facilitating internal operations, like digital meeting platforms.

 The goal of the SAFe transformation under study is to make the entire company operate in an agile manner. The bank has been transforming according to SAFe since 2018 and has a handful Agile Release Trains in operation. The bank chose SAFe as it was considered to have the highest chance to be accepted and trusted by decision-makers. Another reason for selecting SAFe was the standardization of roles and practices, that makes it easier to both educate employees and recruit new ones. The company has previous experience with agile on a smaller scale, and some individual teams apply agile practices in their daily work.

Prior to the interviews, we created a list of categories where the challenges have been previously identified based on the input from literature reviews [1][5][2]. To assure completeness, we ensured that all categories identified by the three systematic literature reviews fit into the proposed categories. On top of that, we added a "Banking specific" category to capture the challenges that are specific to the domain under study. Figure 4 shows the resulting categories. These categories were used as a base for each interview. During the interviews, the interviewees were asked if the category was relevant for their transformation, and to provide examples of challenges they faced.

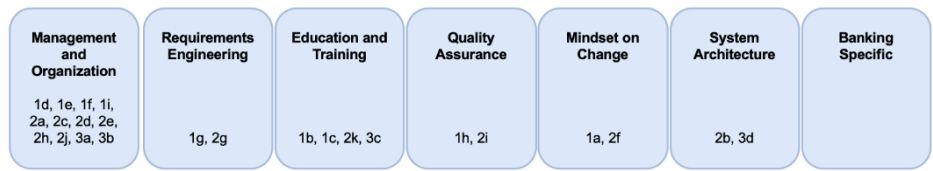

**Fig. 4.** The different categories of challenges in a SAFe transformation used as a foundation in the study. The references relate to the challenges reported in previous secondary studies, see Figure 2.

The interviews were performed with four representatives from the bank. The sampling of the interviewees was motivated by their experience and expertise in the SAFe transformation in the bank. They all belong to the selected group of individuals who were driving the agile transformation in the company. As such, they are expected to have wide knowledge of obstacles and challenges not only in their respective areas, but also in the bank in general. Dealing with such challenges is an important part of their daily work.

A semi-structured interview is common to use in such studies, as such interviews are exploratory and descriptive [17]. The use of interviews with both open-ended and specific questions allows for answers with unforeseen information but still makes sure that the interview stays on topic. Since the respondents have different roles in the transformation, they were not asked the exact same questions. However, each interview followed the same pattern. We started with establishing some background information about the interviewees and their roles in the agile transformation. After that we went through the categories from Fig. 4. For each category the interviewees were asked to provide information about the challenges that the bank faced in the respective area.

To ensure that the responses were interpreted correctly, the transcript was sent to the interviewees for validation. Based on the transcripts, each of the three involved researchers (authors 1-3) performed an independent identification and categorization of the challenges. Later, a workshop was arranged to agree on a joint list of categorized challenges. The final list was compared to the findings from other studies.

## 5 Results

The results are summarized in Table 1 and further elaborated in this chapter.

**Table 1.** The resulting list of challenges

| **Management and organization** |
|---|
| A1. Difficult to adapt existing internal processes to the agile development practices |
| A2. Difficult to define and adapt to new roles |
| A3. Difficult to create a shared vision and align the entire organization around common goals |
| A4. Challenge to make the entire organization to work agile |
| A5. Hard to coordinate and align agile development in a distributed organization |
| **Education and training** |
| B1. Competency gap between old and new roles |
| B2. Difficult to achieve optimal time planning for the training activities |
| B3. Need for tailoring of training to meet different needs in the organization |
| **Culture and mindset** |
| C1. Individual resistance to change |
| C2. Change is discouraged in the banking industry |
| C3. Banking values traditions and stability, implies difficulties for transformation |
| **Requirements engineering** |
| D1. Difficult to prioritize requirements |
| D2. Difficult to break down requirements |
| D3. Hard to find a balance between specificity and time |
| D4. Requirements in SAFe defined differently, new way of dealing with requirements |
| **Quality assurance** |
| E1. Difficult to balance speed and quality |
| E2. Challenge to increase test automation |
| **Systems architecture** |
| F1. Legacy systems are not easily adopted to agile ways of working |
| F2. Complexity and interdependencies between legacy systems are hard to deal with |
| **Banking specific** |
| G1. External rules and regulations complicate transformations in the banking industry |
| G2. Big transformations may adventure the external stakeholder trust |

**Management and organization (A):** In general, software development teams are more likely to have the knowledge or even previous experience with agile methods and the transformation might therefore not be as challenging for them as it is for other departments. The challenge is instead to integrate and engage every unit that are in a large organization in the SAFe transformation, especially if they have traditionally been working in non-agile ways. This challenge is exacerbated when developers are not collocated.

**Education and training (B):** The case company faces a competency gap in profiles of old and new roles as the transformation is taking place. But despite the positive attitude on employees' competency, there are difficulties with bridging the competency gap that the SAFe transformation requires. This highlights that there is need for better training when adopting SAFe. To be able to adopt SAFe successfully, it is important to educate all Scrum masters of the purpose and the principles behind SAFe for them to, in turn, be ambassadors within their teams. However, the case company faced difficulties in providing effective training activities. The main issues relate to the need for tailoring training to the companies' and needs, and also finding optimal time for training and minimizing impact on productivity.

**Culture and mindset (C):** The main challenge of adopting SAFe is that the employees are not used to making changes. It was reported that there is no company infrastructure for supporting change and transformation management, which leads to people not wanting or expecting change. One interviewee mentioned that the banking industry in general values traditions and that many people are used to working in a way where it is easier to predict outcomes and to make long-term planning, in contrast to agile methods. The case company has a successful business, and it is then hard for some employees to understand why a transformation as big as adopting SAFe should be made.

**Requirements engineering (D):** The traditional requirements engineering role has faced one of the most significant changes in the transformation. One interviewee mentioned that it is a challenge to change the requirements process to instead begin with expressing high-level business needs that the development team should build from, rather than begin with a complete list of requirements as in the previous way of working. In addition, rationale for requirement prioritization from upper management is not clearly communicated to developments. The reason for this might be, according to one interviewee, that the SAFe transformation was started at the bottom of the organization.

An interviewee revealed that there is a problem with teams viewing agile epics as just a high-level requirement that should be broken down into smaller requirements when it should instead be viewed as a placeholder for communication. An epic should be tied to a story so that everyone involved understands what is desired to be achieved. By communicating the epic as a story instead of through conventional documentation, the chance of success increases, as stated by the interviewee. In the case company, this has shown to be a long process, taking up to three years before the new way of dealing with requirements has become universally accepted.

**Quality assurance (E):** In a more traditional development process, software testing occurs primarily towards the end of the process. On the other hand, agile methods strive for built-in-quality where testing takes place continuously in the process. One interviewee reported that when changing the way of working with testing, the case company faces several challenges. One reported challenge is building a foundation for automated testing, and another one was the challenge of balancing quality and time.

**Systems architecture (F):** One challenge mentioned in the interviews is that the complexity of the systems makes it difficult to work according to SAFe. Dependencies between systems and having many systems are all impediments for the SAFe transformation. The complexity of the systems makes it harder for the case company to deliver customer value for every sprint (i.e., every two weeks), e.g., compared to a fast-moving small startup. However, according to the interviewees, it will never be relevant to increase the sprint length.

**Banking specific (G):** The case company's reputation is dependent on releasing reliable products and services that people can trust. However, balancing reliability and frequent releases is a delicate issue. The interviewees mentioned the risk of adventuring external stakeholders' trust when working with more frequent releases. Simultaneously, new and sometimes unpredictable regulations and quickly changing demands in the market put pressure on the case company's ability to be flexible. In an industry where more niche banks take market shares, the competitive landscape puts pressure on the large full-service banks to become fast and flexible while not jeopardizing the trust and loyalty of external stakeholders that are more used to stability and a slow pace of changes.

## 6   Discussion

The challenges of adopting SAFe can vary between different banks. We believe that by comparing the study result with other experience reports some common challenges can be identified. All the challenges that were found in the case company are listed in Table 1. We have found that many of them are also mentioned in experience reports from other banks, or even in other industries.

We show that the case company faced several challenges in the "management and organization" category. Some of them are also reported by other banks. Capital One [15] reports dependencies outside of the teams as a challenge, similar to our findings at the case company. This suggests that organizational boundaries in companies affect the SAFe transformation. Here, an interesting dilemma appears. It is a common practice not to deploy the transformation all at once but instead introduce changes incrementally, as mentioned in the cases of ING Benelux, BNP Paribas, Deutsche Bank, and SimCorp [12]. At the same time, dependencies between departments are an issue for a successful transformation, implying that changes need to occur at many departments at once. Not surprisingly, the challenge to transform the entire organization into an agile way of working is not unique to the banking industry. Ericsson also reports a challenge with surrounding organizations being in "waterfall mode" when scaling the agile methods [9]. Furthermore, the challenge of redefining roles is also reported at Standard Bank [13], confirming that is evident across banks transforming according to SAFe.

Experience reports from other banks also cover challenges identified in the "education and training" category. One of them is the need for tailoring of training to meet different needs in the organization. Standard Bank reports difficulties for uniting teams when scaling up the transformation [13], and Nordea reports a challenge with aligning

11different agile teams [11]. This emphasizes the need to align knowledge and competency about agile practices across all teams. The challenge of aligning agile teams could also be a challenge related to management and organization since a lack of shared vision and optimization targets might be a reason why teams are not synchronized.

For the challenges in the "culture and mindset" category, we show that the general characteristics of the banking industry culture pose particular difficulties for making changes in general. This has also been found in another bank, where accepting change is reported as a challenge in the case of Capital One [15].

There are also some challenges reported in the "requirements engineering" category. One of the challenges is reported by Standard Bank. The report states that having transparent requirements prioritization is difficult [13]. This challenge further shows a need for management to formulate and communicate goals in the teams. Previous research suggests that effective communication can facilitate the prioritization of requirements [18]. The requirements engineering-related challenges in SAFe transformation generalize beyond the banking industry as managing and planning requirements are also mentioned as challenges at Nokia [8].

When it comes to "quality assurance," the issue of balancing quality and speed of delivery found in this study is also found in the experience report from Standard Bank [14]. The challenge of managing more frequent releases is also related the "systems architecture" category, where the complexity of the systems and dependencies between the systems are found challenging for the SAFe transformation in the case company.

The case company faces challenges originating from the external stakeholders, as reported in the "banking specific" category. The company operates in an industry where every product has to undergo rigorous testing procedures before releasing due to external regulations. The testing process can be hard to align with agile methods. Similar issues with agile methods and regulated development are discussed in safety-critical contexts [19]. In the banking domain, the characteristics of the services necessitate instilling high levels of trust, which can be difficult when having to work with incremental updates and frequent releases, as recommended by agile principles.

To summarize, Table 2 maps the challenges identified in our study to other studies and experience reports reporting similar findings. The distinction is made between findings from within and outside of the banking industry. We find that some of the challenges are found in other banks, and that a majority of the challenges is common with SAFe transformations in other industries. The challenges that do not appear in previous customer stories or research are:

- C3. Banking values traditions and stability, implies difficulties for transformation
- G1. External rules and regulations complicate transformations in the banking industry
- G2. Big transformations may adventure the external stakeholder trust

All these challenges are very specific to the banking industry, and can be considered sensitive, which may explain why they have not been reported previously. As these challenges seem not to have been obvious from the beginning, we believe that they are one of the unique contributions of our study and can be of particular interest for other actors in the banking industry.

Table 2. Mapping between challenges found in this study and related work

| Challenge in our study | Challenge found in other banks | Challenge found in other industries |
|---|---|---|
| A1 | [15] | [2] |
| A2 | [13] | [2] |
| A3 |  | [5] |
| A4 |  | [9] |
| A5 |  | [2] |
| B1 |  | [2] |
| B2 |  | [10] |
| B3 | [13] | [9] |
| C1 |  | [5] |
| C2 | [15] | [10] |
| C3 |  |  |
| D1 | [13] | [2] |
| D2 |  | [5] |
| D3 |  | [5] |
| D4 |  | [5] |
| E1 | [14] | [5] |
| E2 |  | [2] |
| F1 |  | [2] |
| F2 |  | [5] |
| G1 |  |  |
| G2 |  |  |

## 7   Threats to Validity

This section discusses threats to the validity of our conclusions. The discussion is organized into construct validity, external validity, and reliability issues. We do not discuss internal validity, as our conclusions include no causal claims.

Construct validity reflects how well the phenomenon under study is captured. The researchers have substantial pre-understanding of agile transformations and combined experience of almost two decades of large-scale agile transformations at five different companies. Regarding the constructs under study, we rely on standard SAFe concepts that were well understood by all interviewees. Thus, we consider the threats to construct validity as minimal.

External validity is related to the generalization of the findings outside the studied setting. We claim that our conclusions are relevant for other large banks. To mitigate the threats to external validity, we selected interviewees with broad experience and insights in various units within the bank – some of them also had worked with other banks in the past. We have also identified an overlap between our findings and what other banks undergoing similar transformation report in white papers, as well as an overlap between our findings and the general body of knowledge with respect to agile transformations. We believe that strengthens the generalizability claim. Further studies may reveal that our findings, at least in part, also generalize to smaller challenger banks [20] and other FinTech businesses such as insurance companies.

The reliability of a study is related to the dependence on specific researchers. We mitigate the threats to research bias by applying established research practices. The interview guide was co-developed iteratively by the first three authors. The interviews were conducted, recorded, and transcribed by the first author. Three authors independently analyzed the transcripts, and a joint workshop was organized to summarize the results. The few findings that deviated were discussed until a common understanding was reached. We maintained a chain of evidence from the conclusions to individual interview statements through fine-granular traceability during the study. However, for confidentiality and anonymity reasons, we agreed with the case company and with the interviewees not to reveal exact mappings between statements and interviewees in this paper. Nevertheless, it is possible that another set of researchers would emphasize other aspects of SAFe transformations. However, as the reported findings are presented on a high-level of detail, we consider the threats to the reliability of the study as minor.

## 8    Conclusions

The goal of this study was to identify challenges of a SAFe transformation in the banking industry. To address the research question, we have performed a study at a large full-service bank. In the study, we have identified several challenges belonging to seven categories, ranging from technical challenges related to the system architecture to banking specific issues. Significant challenges include the alignment of goals and optimization targets within the entire organization. Our findings considerably overlap with experience reports from similar transformations, both in the banking industry and in other industries. Consequently, we believe that the findings are interesting for the banking industry and, therefore, are relevant to other banks that are about to embark upon their SAFe transformation journeys. As a natural next step we would like to investigate how the challenges have been addressed at the bank to be able to provide actionable recommendations.

## 9    REFERENCES

[1]    K. Direkt, M. Paasivaara, and C. Lassenius, (2016). "Challenges and success factors for large-scale agile transformations: A systematic literature review", Journal of Systems and Software, vol. 119, pp. 87-108.
[2]    A. Putta, M. Paasivaara, and C. Lassenius, (2018)."Benefits and challenges of adopting the Scaled Agile Framework (SAFe): Preliminary results from a multivocal literature review", In Proc. of the International Conference on Product-Focused Software Process Improvement, pp. 334-351.
[3] Version One, (2020). "14th Annual State of Agile Report", https://stateofagile.com/#ufh-i-615706098-14th-annual-state-of-agile-report/7027494

[4]   Scaled Agile, (n.d.). "SAFe customer stories", retrieved 2020-08-25 from: https://www.scaledagile.com/customer-stories/
[5]   Ö. Uludag, M. Kleehaus, M. Caprano, and F. Matthes, (2018). "Identifying and structuring challenges in large-scale agile development based on a structured literature review", In Proc. of the 22nd International Enterprise Distributed Object Computing Conference, pp. 191-197.
[6]   Scaled Agile (Dec 2019), "Achieving business agility with SAFe® 5.0", https://www.scaledagile.com/?ddownload=47510
[7]   R. Knaster, and D. Leffingwell, (2017). "SAFe 4.0 distilled: applying the Scaled Agile Framework for lean software and systems engineering", Boston: Addison-Wesley Professional.
[8]   M. Laanti, O. Salo, and P. Abrahamsson, (2011). "Agile methods rapidly replacing raditional methods at Nokia: A survey of opinions on agile transformation", Information and Software Technology, vol. 53, pp. 276-290.
[9]   M. Paasivara, B. Behm, C. Lassenius, M. Hallikainen, (2018). Large-scale agile transformation at Ericsson: a case study", Empirical Software Engineering, vol. 23, pp. 2550-2596.
[10]  M. Paasivaara, (2017). "Adopting SAFe to scale agile in a globally distributed organization", In Proc. of the 12th International Conference on Global Software Engineering, pp. 36-40.
[11]  Scaled Agile, (2015). "SAFe Case Study: Nordea", retrieved 2020-08-30 from https://www.scaledagile.com/case_study/nordea/
[12]  Everaerts, S, (2018). "Initio", retrieved 2020-08-30 from: https://www.initio.eu/blognavigation/2018/12/3/embracing-scaled-agile-framework-in-banking-amp-investment-industry
[13]  Scaled Agile, (n.d.). "SAFe Case Study: Standard Bank," retrieved 2020-08-30 from: https://www.scaledagileframework.com/standard-bank-case-study/
[14]  K. A. Johnston, and G. Gill, (2017). "Standard Bank: The Agile Transformation", vol. 6, no. 1.
[15]  Scaled Agile, (n.d.). "SAFe Case Study: Capital One," retrieved 30/8, 2020 from: https://www.scaledagileframework.com/capital-one-case-study/
[16]  A. Berkani, D. Causse, and L. Thomas, (2019). "Triggers analysis of an agile transformation: the case of a central bank", Procedia Computer Science, pp. 449–456.
[17]  P. Runeson, and M. Höst, (2009)." Guidelines for conducting and reporting case study research in software engineering", Empirical Software Engineering, vol. 14 (no. 2), pp. 131-164.
[18]  E. Bjarnason, K. Wnuk, B. Regnell, (2011)." Requirements are slipping through the gaps – A case study on causes & effects of communication gaps in large-scale software development", In Proc. of the 19th International Requirements Engineering Conference, pp. 37-46.
[19]  J.-P. Steghöfer, E. Knauss, J. Horkoff, and R. Wohlrab (2019). Challenges of scaled agile for safety-critical systems. In Proc. of the International Conference on Product-Focused Software Process Improvement, pp. 350-366.
[20]  S. Blakstad and R. Allen, "New standard models for banking," FinTech Revolution, Palgrave Macmillan, Cham., pp. 147-166.